\documentclass[]{aa}

\usepackage{natbib}
\bibpunct{(}{)}{;}{a}{}{,}
\usepackage{txfonts}

\usepackage{graphicx}

\title{On the distance of GRO J1655-40\thanks{Based on data obtained at the Very Large Telescope, European Southern Observatory, under program ID 073.D-0473(A).}}

\author{C. Foellmi\inst{1,2} \and E. Depagne\inst{1,3} \and  T. H. Dall\inst{1} \and I. F. Mirabel\inst{1,4}}
\institute{European Southern Observatory, 3107 Alonso de Cordova, Casilla 19001, Vitacura, Santiago, Chile \and
Laboratoire d'Astrophysique, Observatoire de Grenoble, BP 53, 38041 Grenoble Cedex 9, France \and
Departamento de Astronom\'ia y Astrof\'isica, Pontificia Universidad Cat\'olica de Chile, Campus San Joaqu\'in. Vicu\~na Mackenna 4860 Casilla 306, Santiago 22, Chile.
\and
On leave from Service d'astrophysique-CEA-Saclay, 91191 Gif-sur-Yvette, France
}

\offprints{C.Foellmi \email{cfoellmi@eso.org}}

\abstract
{}
{We challenge the accepted distance of 3.2~kpc of GRO~J1655-40. We present VLT-UVES spectroscopic observations to estimate the absorption toward the source, and determine a maximum distance of GRO~J1655-40.}
{We show that the accepted value of 3.2~kpc is taken for granted by many authors. We retrieved in the ESO archive UVES spectra taken in April 2004 when GRO~J1655-40 was in quiescence to determine the spectral type of the secondary star. For the first time we build a flux-calibrated mean (UVES) spectrum of GRO~J1655-40 and compare its observed flux to that of five nearby stars of similar spectral types. We strengthen our results with the traditional pair method, using published photometric data of the comparison stars.}
{We show that the distance of 3.2~kpc is questionable. We determine a spectral type F6IV for the secondary star. We demonstrate in details that the distance of GRO~J1655-40 must be smaller than 1.7~kpc.}
{The runaway black hole GRO~J1655-40 could be associated with the open cluster NGC~6242 which is located at 1.0$\pm$0.1~kpc from the Sun. At $D \leq$ 1.7~kpc the jets are not a superluminal, and GRO~J1655-40 becomes one of the closest known black holes to the Sun.}


\date{Received $<$date$>$, Accepted $<$date$>$}

\titlerunning{On the distance of GRO J1655-40}

\begin{document}

\maketitle

\section{Introduction}

\object{GRO~J1655-40} (a.k.a Nova Sco 94) is a Soft X-ray transient (SXT) in our Galaxy (l=345.0$^\circ$, b=2.2$^\circ$; R.A.=16$^h$54$^m$00$^s$, Dec.=$-39^\circ$50$^m$45$^s$). It has been discovered on July 27, 1994 with \emph{BATSE} on board the Compton Gamma Ray Observatory \citep{Zhang-etal-1994}. Optical photometry of GRO~J1655-40 revealed a double-waved modulation with a period of 2.6 days \citep{Bailyn-etal-1995b,vanderHooft-etal-1997}. Strong evidence that the compact object in GRO~J1655-40 is a black hole was presented by \citet{Bailyn-etal-1995b} who established a spectroscopic period of 2.601 $\pm$ 0.027 days. \citet{Shahbaz-etal-1999} published the masses of the black hole and the secondary star: 5.5--7.9 $M_{\odot}$ and 1.7--3.3 $M_{\odot}$ respectively. Given a distance of about 3~kpc, the jets appeared to be superluminal: 1.5$\pm$0.4~$c$ \citep{Tingay-etal-1995}, 1.05~$c$ \citep{Hjellming-Rupen-1995}, where $c$ is the speed of light. Various spectral types of the secondary star have been published. We mention in particular: F3-F6~IV \citep{Orosz-Bailyn-1997}, and F7IV-F6III \citep{Israelian-etal-1999}. The distance of GRO~J1655-40 that is used in the literature is 3.2$\pm$0.2~kpc, although \citet{Mirabel-etal-2002} pointed out that a distance of about 1~kpc cannot be ruled out with the current data. The aim of this paper is to show that none of the distance estimate methods used for GRO~J1655-40 is good enough to claim a distance of 3.2~kpc, and that this distance is actually based on one very uncertain interpretation. We present unpublished optical data of GRO~J1655-40 to show that the distance of the source is certainly smaller than 1.7~kpc.

\section{The problem of the distance}

The distance of GRO~J1655-40 usually quoted in the literature is the one determined by \citet{Hjellming-Rupen-1995} who present new radio data obtained with the \emph{VLA} and \emph{VLBA}. The authors build a kinematic model of the radio jets of GRO~J1655-40 and obtain a distance of 3.2$\pm$0.2~kpc. The authors do not actually measure the distance directly, but refer to  \citet{McKay-Kesteven-1994}, \citet{Tingay-etal-1995} and \citet{Harmon-etal-1995}. Their model provides a refinement of the distance range obtained by the cited authors. \citet{Harmon-etal-1995} rely as well on \citet{McKay-Kesteven-1994} and \citet{Tingay-etal-1995} for the distance. 

The paper by \citet{McKay-Kesteven-1994} simply states: "HI observations of GRO J1655-40 made with the \emph{AT Compact Array} show solid absorption in the velocity range $+10$ to $-30$ km~s$^{-1}$, with a further isolated weak feature at $-50$ km~s$^{-1}$. The balance of probabilities is that the distance is around 3.5~kpc, unless the $-50$ km~s$^{-1}$ feature is due to an atypical cloud." Although providing important information on the absorbing material in the direction of GRO~J1655-40, this IAU Circular is not a measurement of the distance.

Independently, \citet{Tingay-etal-1995} presented new $VLBI$ and $ATCA$ data of GRO~J1655-40. Their HI spectrum obtained with \emph{ATCA} (see their Fig.~2) shows a multi-component profile, with a weak feature at $-50$~km~s$^{-1}$ too. They intended to interpret the various components in terms of HI clouds and HII regions participating to the mean Galactic rotation, and concluded that a minimum distance of 3.0~kpc can be inferred. They finally give a range of 3 to 5~kpc possible for GRO~J1655-40. This estimated distance range is relying on the assumption that the weak feature observed at $-50$~km~s$^{-1}$ is not moving with a peculiar velocity. Although it looks reasonable, this assumption might simply not be true. These measurements are very dependent on the distribution and velocities of various HI clouds along the line of sight; a difficulty that has been claimed to be important by \citet{Mirabel-etal-2002} who note that, in this direction, there are clouds with anomalous velocities up to $-50$ km~s$^{-1}$ \citep{Crawford-etal-1989}, i.e. with an amplitude similar to that of the weak feature observed in the \emph{ATCA} spectrum. 

Moreover, \citet{Mirabel-etal-2002} have shown that the above radio data of \citet{Hjellming-Rupen-1995} allow to derive with certainty only a relativistic {\it upper limit} of $D \leq 3.5$~kpc, since the relativistic time delay of the motion of the ejecta in the sky is given by two equations with three unknowns: the angle with the line of sight of the jet axis $\theta$, the velocity of the jet $\beta = v/c$ and the distance $D$. \citet{Hjellming-Rupen-1995} obtained a distance of 3.2~kpc by assuming $\theta \sim 84^{\circ}$, which actually corresponds only to the upper limit allowed for GRO~J1655-40 by the above equations \citep[see also][]{Mirabel-Rodriguez-1994}. However, \citet{Mirabel-etal-2002} claimed that the assumption that the jet axis is parallel to the axis of the orbital plane (i.e. to the contrary of Hjellming \& Rupen) is equally consistent with the observations at radio wavelengths. 

A distance of $\sim$3.0~kpc was also proposed on the basis of optical data by \citet{Bailyn-etal-1995a}. The method used is the "D" method, following \citet{Jonker-Nelemans-2004}. They have first measured the equivalent widths (EW) of NaI-D lines, from which they claim a color excess $E(B-V)$ of 1.15. They used the results of \citet{Barbon-etal-1990}, who studied the type-Ia supernova 1989B in \object{NGC 3627}. These latter authors mention that the color excess can be properly derived by studying the detailed structure of the interstellar CaII and NaI-D lines, but to do so, a high-resolution and high Signal-to-Noise ratio spectrum is needed, which is not their case nor the one of the spectrum of \citet{Bailyn-etal-1995a} that has a resolution of $\sim10$\AA. Consequently, \citet{Barbon-etal-1990} determined an empirical and roughly linear relation between the equivalent width of the NaI-D lines and the color excess $E(B-V)$ {\it for six supernovae}. Although not precisely stated, \citet{Bailyn-etal-1995a} must have used this empirical relation to determine their value of the color excess, and claimed that the results is consistent with the EW of other interstellar lines in the optical domain. To obtain the distance, they use the standard relations between the absorption, the reddening, the colors and the apparent and absolute magnitudes \citep[see e.g.][]{Allen-1973}. Finally, they conclude that the distance of the source is compatible with $D \sim 3$~kpc, "in agreement with the radio observations" of \citet{Tingay-etal-1995}.

We note that their result is based on the assumption that the absorbing material is distributed homogeneously between the source and the observer. This has been again challenged by \citet{Mirabel-etal-2002} who mention in particular that the reddening in this region of the sky occurs in the local arm within 700~pc from the Sun.

Similarly, \citet{Bianchini-etal-1997} measured the color excess also by measuring the equivalent widths through gaussian fitting of the interstellar 5980\AA\ NaI-D doublet and the 6613\AA\ line in their "higher resolution" data. They use the relationships between equivalent widths and color excess given by \citet{Herbig-1975} and \citet{dellaValle-Duerbeck-1993} to obtain a (mean) color excess of 1.13, in agreement with the value found by \citet{Bailyn-etal-1995a}. With a resolving power of about 3700, not only their spectrum cannot resolve the multi-component profile of the NaD lines, but also miss the fact that the lines appear saturated (see our Fig.~\ref{sodium}), and therefore, simply cannot be used to measure the extinction.

On the other hand, \citet{Greiner-etal-1995} presented new \emph{ROSAT} X-ray data, from which they infer a distance of "about 3~kpc". Their method consists of fitting the halo of the observed radial profile of the source, produced by the scattering of the X-rays by the interstellar dust. To fit the radial profile of GRO~J1655-40 they assume an uniform dust distribution between the observer and the source. From this fit, they obtain, a value of the effective optical depth at 1~keV of $\tau_\mathrm{eff} \sim 0.3$.

Furthermore, they used the results of \citet{Predehl-Schmitt-1995} who have studied in details X-ray halos in \emph{ROSAT} sources. From the fractional halo intensity it is possible to derive the dust column density. The authors show that a good correlation exists between the simultaneously measured dust and hydrogen column densities, indicating that gas and dust must be to a large extent cospatial. From this correlation that depends on the optical depth $\tau_\mathrm{eff}$(1~keV), \citet{Greiner-etal-1995} obtain for GRO~J1655-40 an absorption of $A_V = 5.6$ mag and a hydrogen column density of $N_H = 7.0 \times 10^{21}$ cm$^{-2}$. However, as emphasized by \citet{Jonker-Nelemans-2004}, their implicit assumption is that the sight-line for GRO~J1655-40 has the same gas-to-dust ratio as the sight-lines for which the relations between $A_V$ and $N_H$ have been established. To finally compute the distance, \citet{Greiner-etal-1995} use the mean extinction law given by \citet{Allen-1973}: $A_V = 1.9$ mag~kpc$^{-1}$. As noted by \citet{Mirabel-etal-2002}, GRO~J1655-40 is in the Scorpius region of the sky which contains rather clumpy optical dark clouds in the foreground. Thus the mean extinction law might not be accurate enough in this case.
 
Similary, \citet{Orosz-Bailyn-1997} could have used the "B" method of distance determination \citep{Jonker-Nelemans-2004}, but they {\it assumed} a distance of 3.2~kpc again citing \citet{Hjellming-Rupen-1995}, and a color excess of $E(B-V) = 1.3 \pm 0.1$; this latter value being obtained by \citet{Horne-etal-1996} who used high-quality UV spectra obtained with $HST$. However, the cited paper of \citet{Horne-etal-1996} is an IAU circular where it is simply stated that "deep 220-nm absorption in the HST spectrum {\it suggests} $E(B-V)$ = 1.3 mag."

\citet{Hynes-etal-1998} presented an multi-wavelength dataset, obtained with $HST$, $AAT$, $RXTE$ and $CGRO$. The distance is quoted from \citet{Hjellming-Rupen-1995}. The authors discuss carefully the problem of the extinction. They have measured the reddening of the source by using a power-law fit to the UV data. As they mention, "no other assumptions about the properties of GRO~J1655-40 are required, although [they] do need to assume an extinction curve." They used the mean Galactic extinction curve of \citet{Seaton-1979} and the $R_v$-dependent curves of \citet{Cardelli-etal-1989}. They claim that both curves give identical results, while using the unusual extinction toward $\sigma$ Sco, 17$^\circ$ away from GRO~J1655-40, gives a much poorer fit, without actually mentioning how unusual this extinction is. Their final value is $E(B-V) = 1.2 \pm 0.07$ mag. However, there is again {\it a priori} no reason to think that the mean Galactic extinction curve is valid in that particular direction.

The authors also computed the absolute magnitude of the secondary star. They follow \citet{Orosz-Bailyn-1997} for the spectral type of the secondary (F5IV). They rescale the magnitude to the effective radius of GRO~J1655-40, and obtain an absolute magnitude of $M_V = 0.7 \pm 0.5$, which is very different from the absolute magnitude of an F5IV star as given by, e.g., \citet{Gray-1992}: 3.2 mag. In particular, it is not clear if they assumed that the star was filling its Roche-lobe or not. If true, it is rather straightforward to compute the radius \citep[see e.g.][]{Paczynski-1971,Jonker-Nelemans-2004}. Then, \citet{Hynes-etal-1998} corrected the apparent magnitude (17.12) of \citet{Orosz-Bailyn-1997} to $17.18 \pm 0.06$. {\it Adopting} the distance of 3.2~kpc by \citet{Hjellming-Rupen-1995}, and an average reddening law ($R_V = 3.2$), they obtain E(B-V) $= 1.25 \pm 0.17$ in agreement with previous publications.

\citet{Beer-Podsiadlowski-2002} reanalyzed the lightcurve in quiescence obtained by \citet{Orosz-Bailyn-1997}. In their models, the distance is said to be a free parameter, although it is not clear if this parameter has been allowed to go as low as 1.0~kpc. Moreover, in many places they claim that the distance of 3.2~kpc of \citet{Hjellming-Rupen-1995} has been used "to tighten" their results. They also mention that they obtain a reasonable model of the lightcurves with a rather cool disc, with a distance "much closer than the distance of 3.2~kpc" (and with a higher $\chi^2$). For that reason (i.e. the distance is too small compared to 3.2~kpc), they discarded the model.

The authors emphasize that their model is very dependent on the extinction curve. They note that if the extinction curve were incorrect, the fitting procedure would compensate by choosing different values for the distance, temperature of the secondary star and the color excess $E(B-V)$. Although the temperature must be consistent with the spectral type of the secondary, the distance and extinction could both be wrong and compensate each other. The authors claim to obtain values for $D$ and $E(B-V)$ in agreement with the (possibly incorrect) 3.2~kpc distance and (again) a mean Galactic extinction curve. However, even reasonable, a such mean extinction curve cannot be used to strengthen the confidence of the results.

\section{New observational material}

Our dataset on GRO~J1655-40 consists of VLT-UVES spectra available in the ESO archive\footnote{{\tt http://archive.eso.org}.} (prog. ID 073.D-0473(A), P.I. Rebolo). The spectra were obtained in 2004 with three different central wavelengths: 4025\AA\ on the UVES blue side, 5265\AA\ comprising H$\beta$ and 6300\AA\ comprising H$\alpha$ on the UVES red side. The observations were carried out on April 16, 18, 19, 22 and 25, and also June 18 of 2004. Apart from April 22, two spectra of 1440 seconds were obtained for each of the three central wavelengths. For every setting, we have retrieved in the the archive all possible calibrations: biases, wavelength calibration frames, flat-fields, order definition frames and format check frames. 

All the spectra have been reduced using the UVES pipeline \citep[see e.g.][]{Ballester-Hensberge-1995}. The reduction process includes bias and inter-order background subtraction (object and flat-field), optimal extraction of the object (above sky, removing sky lines and rejecting cosmic ray hits), division by a flat-field frame extracted with the same weighted profile as the object, wavelength calibration, rebinning to a constant wavelength step and merging of all overlapping orders. The spectra have been shifted to the heliocentric restframe, and smoothed with a boxcar of 10. We emphasize that the sky line subtraction performed by the pipeline is made through the detection of a line that cross the whole slit and is producing a flat contribution to the flux across the slit. This important point is discussed below. 

For the flux calibration, we have used the method described in the UVES quality control pages\footnote{See {\tt http://www.eso.org/instruments/uves/}}. This consists of normalising each reduced spectrum by the exposure time, correcting for the gain of each CCD and correcting for the atmospheric transmission\footnote{See {\tt \small http://www.eso.org/observing/dfo/quality/UVES/\\qc/response.html} for details.}. The final step consists of taking the UVES master response curve to provide the F-$\lambda$ flux-calibrated and (atmospheric) extinction-corrected spectra. We used the latest master response curve, obtained in September 28, 2004. It  provides a flux calibration with a relative error less than 10\% (see the indicated webpage).

\begin{figure}
\centering
\includegraphics[width=0.9\linewidth]{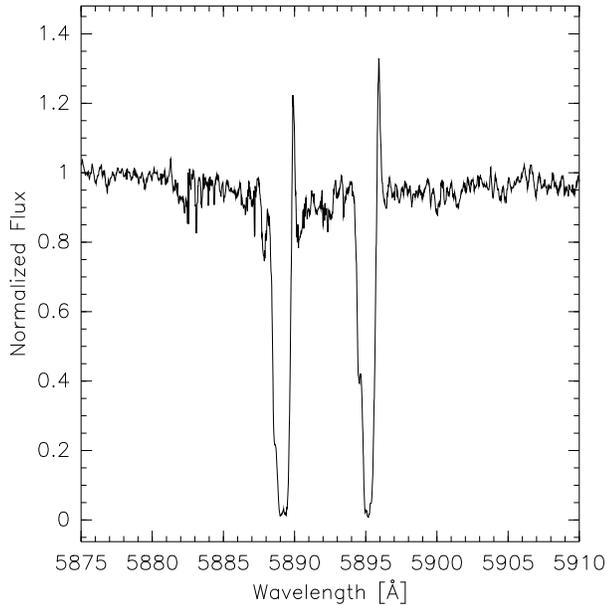}
\caption{Mean UVES spectrum around the saturated NaI-D lines.}
\label{sodium}
\end{figure}

Since the Signal-to-Noise ratio (S/N) of the blue spectra (with a central wavelength of 4025\AA) was very close to unity, they have been discarded from our analysis. The individual red "H$\beta$" and "H$\alpha$" spectra have a S/N ratio between 7 and 20 (before smoothing). They respectively cover the wavelength range from 4785 to 5755\AA\ and 5835 to 6805\AA, both with a dispersion of 0.014740 \AA/pix. The resolving power of all spectra is 45~000. This is, in this case, good enough to resolve the NaI-D lines, that appear saturated (see Fig.~\ref{sodium}).

According to \citet{Brocksopp-etal-2005-astroph}, GRO~J1655-40 remained in quiescence since its discovery outburst by BATSE in July 1994 until 2005 February 17, when an increase in X-ray flux was detected by $RXTE$. Therefore, we can expect that our UVES spectra of GRO~J1655-40 were taken when the source was in quiescence. We show below that this is indeed the case.

\section{Analysis}

\subsection{Spectral separation}

In order to build a clean mean spectrum of the secondary star of GRO~J1655-40 and determine its spectral type, we fitted the continuum of individual spectra and attempted a separation using the iterative reconstruction method described in \citet{Demers-etal-2002}, further enhanced and applied to a Wolf-Rayet+O binary by \citet{Foellmi-etal-2006a}. The reader is referred to the latter paper for a detailed description of the method. It has been in this case applied individually to the serie of spectra around H$\beta$ and H$\alpha$, and allow us to not only "clean" the spectrum of the source by removing any features unrelated to the companion (i.e. a possible contribution of a disk, and/or features that may have been left by the 2D background subtraction of the pipeline), but also to build a correct mean spectrum by aligning the lines of the moving secondary along the orbital motion.

We started by roughly measuring the positions of the center of the H$\beta$/H$\alpha$ absorption profile, and transformed them into radial-velocities (RVs). We used these RVs to shift the spectra, and build a high-S/N absorption-component spectrum. The result is a good first-guess spectrum of the stellar object that produces these absorption lines (that is, the secondary), whereas all unrelated features are smeared out (as much as the RV differences are large). It is then shifted back to the original positions and subtracted from the original spectra. This provides a first-guess of the "second-component" spectrum, clearly revealing narrow emission peaks in all individual spectra. 

\begin{figure}
\centering
\includegraphics[width=0.9\linewidth]{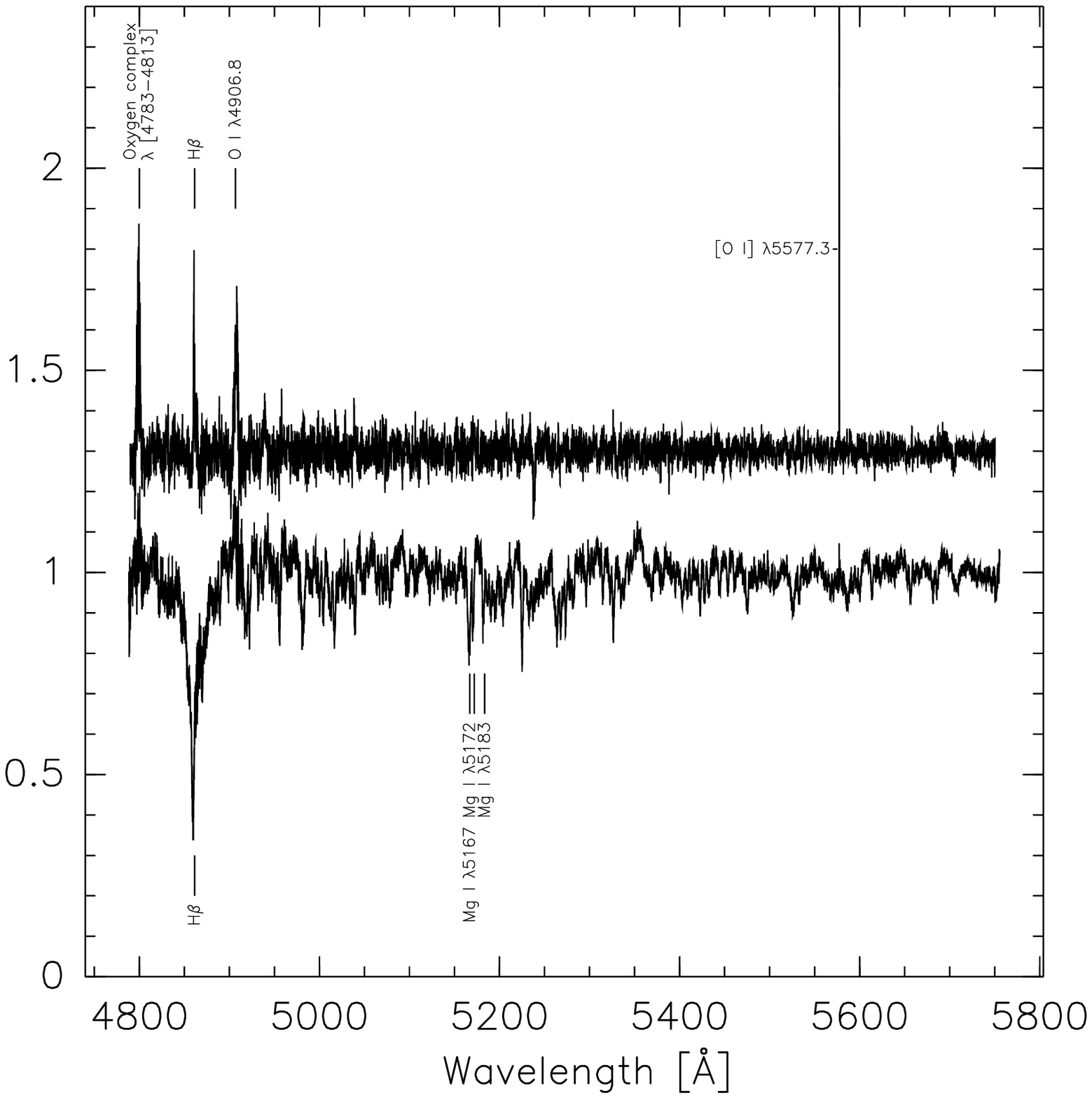}\\
\includegraphics[width=0.9\linewidth]{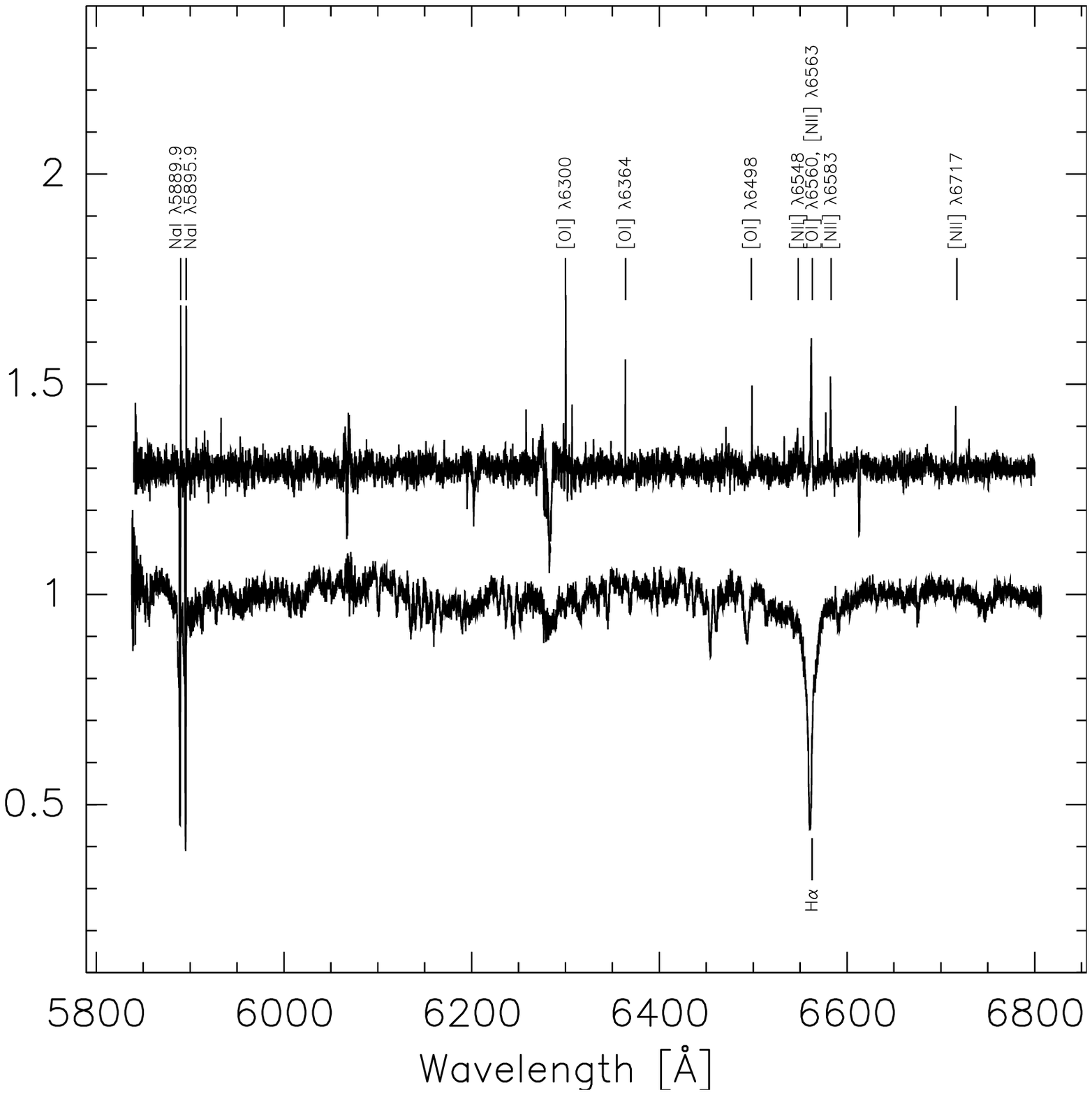}
\caption{Results from the spectral separation procedure. On both panels, respectively showing the H$\beta$ and the H$\alpha$ spectra, the spectrum of GRO~J1655-40 is the lower one, while the extracted sky spectrum is shifted up by 0.3 continuum unit. The mean S/N ratio in the continuum is about 60 to 70. Important lines are labeled.}
\label{extracted_mean}
\end{figure}

We used a cross-correlation (CC) technique to measure the position of the emission peak that is on top of H$\beta$/H$\alpha$, and continued the extraction process. We also used CC for measuring RVs on the extracted absorption-component spectra during the next iterations. We performed in total 3 iterations, since no noticeable changes were visible on the results of subsequent ones. The resulting mean extracted spectra are shown in Fig.~\ref{extracted_mean}.

With the RVs of the absorption lines obtained via the CC measurements, we tried to compute an orbital solution with the period given by \citet[][ P=2.62157 days]{Orosz-Bailyn-1997}. We obtained a solution with an orbital $K$ amplitude of $225\pm10$ km~s$^{-1}$ in excellent agreement with that of Orosz \& Bailyn ($228.2\pm2.2$ km~s$^{-1}$). Our systemic velocity ($-88\pm5$ km~s$^{-1}$) is however not in agreement with them ($-142.4\pm1.6$ km~s$^{-1}$). Since we have very few points and we did not calibrate in an absolute manner our RVs, our value is of course less reliable.

The RVs of the narrow emission peaks found in the H$\alpha$ and H$\beta$ extracted spectra show no RV variations at all within errors (i.e. $< 5$ km~s$^{-1}$). Some of these emission lines clearly have a multi-component profile. It is very likely that these lines are either faint and narrow sky lines missed by the UVES pipeline, or formed by some "warm interstellar medium" (WIM) \citep[see e.g.][]{Reynolds-1985,Domgorgen-Mathis-1994,Sembach-etal-2000}. We are confident that they do not belong to the GRO~J1655-40 system, because of their constancy in RVs, and since their (main) velocity is never consistent with the systemic velocity of GRO~J1655-40. In particular, we identified the lines [O~I]~$\lambda$5577, $\lambda$6300, $\lambda$6364, [O~II]~$\lambda$6498, $\lambda$6560, [N~II]~$\lambda$6548, $\lambda$6563, $\lambda$6583, $\lambda$6717 (see Fig.~\ref{extracted_mean}). We also visually checked that these lines are actually seen in the 2D images of our spectra. Not only do all of them clearly cross the whole slit length, but they also appear too faint and noisy to be caught by the pipeline.

We found no evidences in our spectrum of any lines moving in anti-phase, nor lines that could not be explained by the sky or WIM. In other words, the contribution from a disk around the black hole, if any, is not seen in our spectra. 

\subsection{The spectral type of the companion star}

In order to study the spectrum of the secondary star in GRO~J1655-40, we first performed a spectral synthesis of the H$\alpha$ region. For that we used the UVES POP database of spectra \citep{uvespop}\footnote{{\tt http://www.sc.eso.org/santiago/uvespop/}}, taking advantage of the fact that this provides spectra taken with the same instrument and central wavelength. We used the slowly rotating star \object{HD 156098} as a template \citep[F6IV, V=5.537 mag, T$_\mathrm{eff}$=6480 $K$, $\log g$=3.94, Fe/H=0.09;][]{Edvardsson-etal-1993}, which corresponds almost exactly to the model used by \citet{Israelian-etal-1999}. The synthesis and spectral subtraction was done using STARMOD \citep{Barden-1985,Montes-etal-1995,Montes-etal-2000}. 

We obtained a very good match to the spectrum of GRO~J1655-40, as shown in Fig.~\ref{match}. A rapid check with other F-star templates (see below) confirm that the best match is done with HD~156098, in excellent agreement with the F6III-F7IV spectral types of \citet{Shahbaz-etal-1999}, although we cannot exclude an uncertainty of at most one subclass. This uncertainty has no significant consequences, as already emphasized by \citet{Israelian-etal-1999} in their abundance analysis of GRO~J1655-40.

There is no sure indication (within the errors) of emission fill-in of H$\alpha$ (the equivalent width of the "emission" in the residual of H$\alpha$ is 0.1 \AA). Since we used an averaged spectrum, we could expect to see a overall smeared out (over the  
line) contribution from any accretion disk. This is not the case, and we can conclude that the disk has a negligible contribution (in emission) and that our quiescence spectra are completely dominated by the secondary. In our best fit, we obtain a rotational velocity $v \sin i = 94 \pm 8$ km~s$^{-1}$, in excellent agreement with that of \citet[][ $v \sin i = 93\pm3$ km~s$^{-1}$]{Israelian-etal-1999}. 

Therefore, we conclude that the secondary star of GRO~J1655-40 shows a spectral type F6IV, as observed in our 2004 data.

\begin{figure}
\centering
\includegraphics[width=\linewidth]{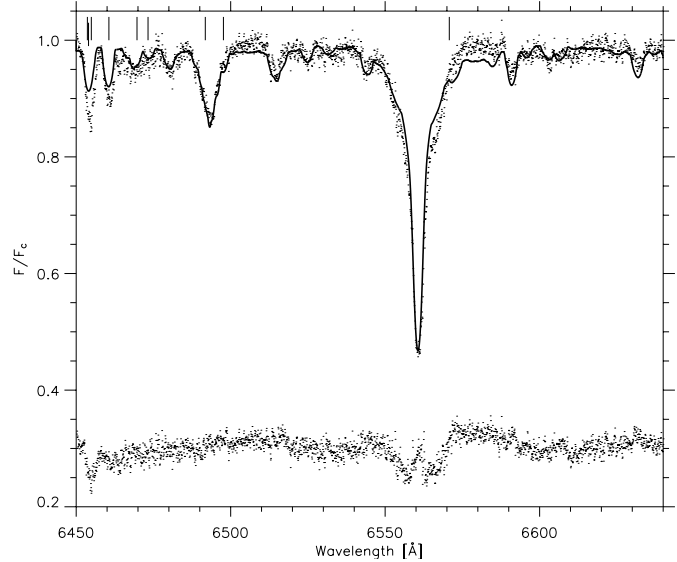}
\caption{UVES spectrum of GRO~J1655-40 (dotted line) with the best fitting synthesized spectrum using a F6IV stellar template broadened to $v \sin i = 94$ km~s$^{-1}$ (solid line). Below are plotted the residuals after subtracting the template from the GRO~J1655-40 spectrum. The residuals are on the same scale, but have been offset by 0.3 for display purposes. Vertical lines denote positions of the Ca lines, and one O line (shifted down).}
\label{match}
\end{figure}

\subsection{Obtaining the distance from the UVES spectra}

Our approach to determine the distance of GRO~J1655-40 is based on the comparison of the flux of GRO~J1655-40 in quiescence with that of a nearby single star of the same spectral type. This is basically the method "A", following \citet{Jonker-Nelemans-2004}. Since there is an uncertainty on this spectral type, we also explore the results with different spectral types of the comparison star.  We strengthen our results by using published photometric data of the comparison stars.

We have flux-calibrated our UVES spectra of GRO~J1655-40 following two methods: using the UVES response curve and using spectrophotometric standard stars observed during the same nights. The comparison between the two methods is relevant because the first method is theoretically not producing an absolute flux-calibration \citep[see e.g.][]{uvespop}. Since we will compare the flux of GRO~J1655-40 with that of F stars who have been calibrated with one of the methods only (namely the method of the master response curve), we must ensure that both methods are equivalent.\footnote{Technically speaking, it seems not necessary to compare the two methods, as long as we use the same master response curve for both GRO~J1655-40 and the comparison F stars. However, it gives a much stronger confidence to the results if we can show that the flux calibration of GRO~J1655-40 is actually good.}

\begin{figure}
\centering
\includegraphics[width=0.9\linewidth]{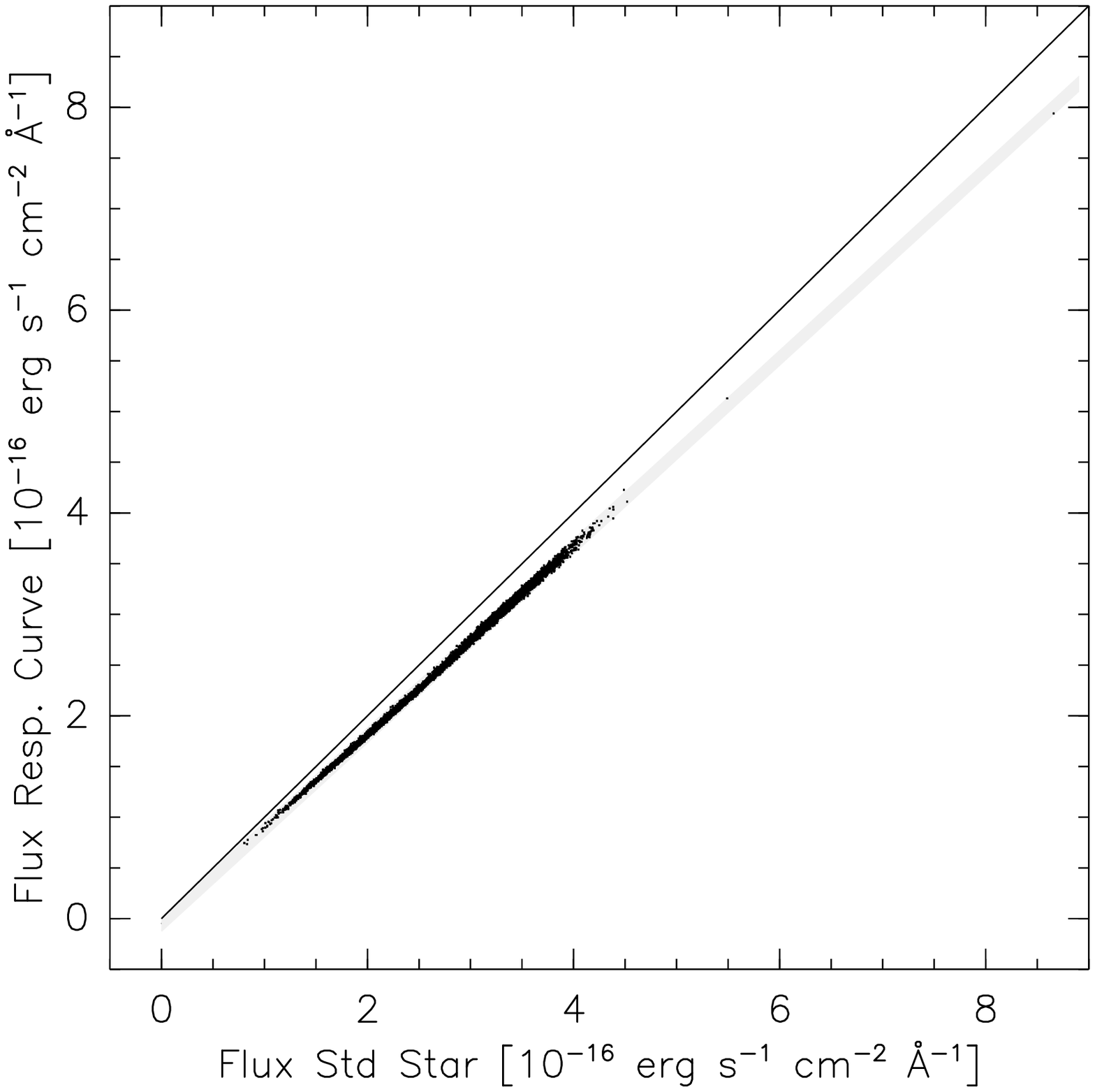}\\
\includegraphics[width=0.9\linewidth]{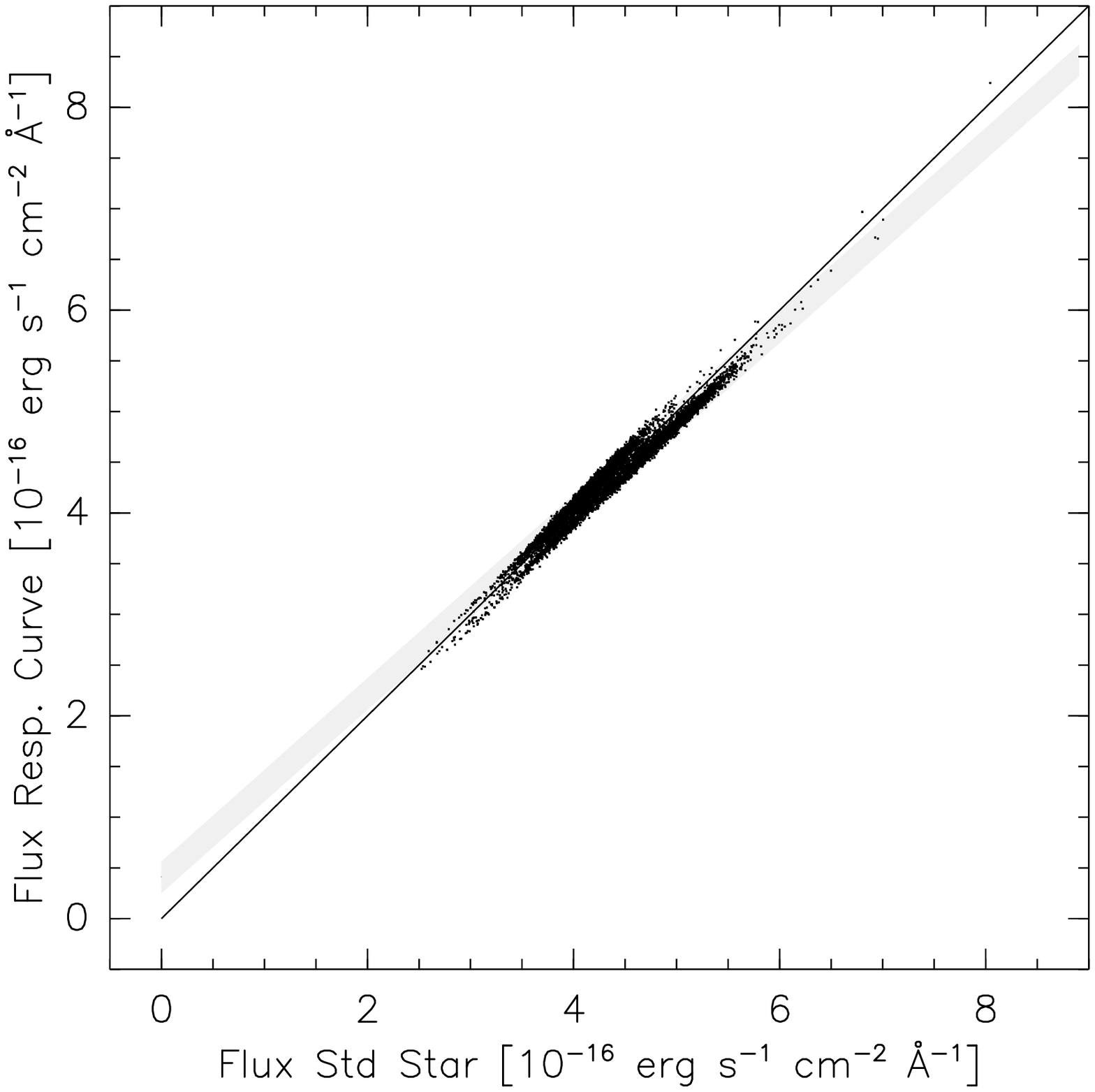}
\caption{Comparison between the two calibration methods. The flux of the mean UVES spectrum calibrated with standard stars is indicated in abscissa, and the flux of the mean UVES spectrum calibrated with response curves indicates in ordinate. Top panel: spectrum with central wavelength of 5265\AA\ (H$\beta$). Bottom panel: spectrum with central wavelength of 6300\AA\ (H$\alpha$). The dark line is bissectrix. The gray area shows the 1-$\sigma$ region around the fitted slope through the points.}
\label{calib-std}
\end{figure}

The first calibration method uses the master response curves, and is described above. The second method was to retrieve observations of standard stars with identical setups from the archive, taken the same nights of observations of GRO~J1655-40. These observations were available only for the nights of April 16 and 18, 2004, and the spectrophotometric standard star used is LTT~3218. To make the comparison between the two methods, we used the spectra of these two nights only. We might argue that slit losses make the comparison pointless. However, it occurred that the seeing was significantly smaller than the slit width (1.6$^{\prime\prime}$ for the observation of GRO~J1655-40 and 5.0$^{\prime\prime}$ for the spectrophotometric standard stars) for the two nights of April 16 (seeing $\sim$0.7) and 18 (seeing $\sim$1.1)\footnote{The values are obtained from the ESO archive DIMM seeing values, available at the time of observations.}. We are therefore confident that the calibration with standard stars is reliable. We finally corrected the spectra for the airmass and the (tabulated) atmospheric extinction. Fig.~\ref{calib-std} shows the flux of the UVES spectrum calibrated with response curve, versus the flux of the mean UVES spectrum calibrated with standard stars.

The difference between the two calibrations are roughly within one $\sigma$ for the so-called H$\alpha$ spectra only, whereas it is not the case for the H$\beta$ spectra. When dividing the two calibrated spectra, for each of the H$\beta$ and H$\alpha$ sets, we obtain the respective ratios: $1.10 \pm 0.01$ and $1.00 \pm 0.02$. We note that the difference between the two methods observed in the "H$\beta$" region is known, as seen in the UVES response curve webpage. We thus decided to pursue the analysis with the H$\alpha$ spectra only. 

In order to determine the amount of absorption towards GRO~J1655-40, we searched for an F6IV star with a known distance. For that matter we used again the spectrum of HD~156098 that has the crucial advantage of having been obtained with the same instrument, and central wavelength. The spectrum has been flux-calibrated with the UVES master response curve. Moreover, this star has a known HIPPARCOS parallax: 19.80$\pm$0.72 mas \citep{hipparcos}. Since its apparent magnitude is known \citep[5.537, ][]{Terzan-Bernard-1981,hipparcos}, we can compute its absolute magnitude: 2.0$\pm$0.2 mag. 

We rebinned the spectrum of HD~156098 to the exact dispersion of that of the H$\alpha$ flux-calibrated spectrum of GRO~J1655-40. We then divided the two spectra and obtained the following mean ratio $f = (8.5 \pm 1.2) \times 10^{-4}$. We make the assumption that the flux from HD~156098 is not affected by extinction, which is a reasonable hypothesis given its proximity (50$\pm$0.2 pc). Therefore, the ratio $f$ of the flux of the F6IV star in GRO~J1655-40 over that of HD~156098 is different from unity because of only the distance and the extinction toward GRO~J1655-40.

We can therefore use the following system of equations:
\begin{eqnarray}
m_1 = -2.5 \log( F_1 ) = & M_1 + a + 5 \log\left( D_1 \right) - 5\\
m_2 = -2.5 \log( F_2 ) = & M_2 + 5 \log\left( D_2 \right) - 5
\end{eqnarray}

where $m_1$ and $m_2$ are the apparent magnitude of GRO~J1655-40 at quiescence and HD156098 respectively, $a$ is the interstellar extinction towards GRO~J1655-40, $M_1$ and $M_2$ the respective absolute magnitudes, and $D_1$ and $D_2$ the respective distances. $F_1$ and $F_2$ indicate the observed flux of GRO~J1655-40 and HD~156098 respectively, and whose ratio is $f$. The difference of the two equations gives:
\begin{eqnarray}
m_2 - m_1 & = & 5\log D_2 - 5\log D_1 + M_2 - M_1 - a \\
          & =  &\frac{5}{2} \log \left( \frac{F_1}{F_2} \right)
\end{eqnarray}

Solving for $a$ we obtain:
\begin{equation}
a = 5 \log \left( \frac{D_2}{D_1} \frac{1}{\sqrt{f}} \right) + M_2 - M_1 \geq 0
\label{absorption_vs_distance}
\end{equation}

This equation can be used to determine a maximum distance of the source. Since we have no precise way to determine the absolute magnitude of GRO~J1655-40 in quiescence, we make the hypothesis that it has the same absolute magnitude to that of the comparison star, {\it more or less one magnitude}, i.e. $M_1 = M_2 \pm 1$. We note that $a$ does not indicate the standard and usual (photometric) absorption in the $V$ band, but rather the absorption within the wavelength limits provided by the spectra. The above equation is consistent as long as we stay within the wavelength limits given by the spectra. These (narrow) limits also allow us to assume that the absorption is constant within the wavelength range. 

In order to see the dependence of the results with the spectral type, we made these calculations for five different F stars with no known peculiarities (such as a binary or variable status), all retrieved from the UVES POP (and used above for spectral synthesis). The five F stars are summarized in Table~\ref{multistar}, where we indicates the star's name, its spectral type, parallax distance and absolute magnitude. The ratio $f$ between the spectrum of GRO~J1655-40 and the spectrum of the F star is indicated. We are confident that these ratios can be considered as constant throughout the wavelength range, since slopes of the ratio spectra were extremely small (comprised between $-1 \times 10^{-7}$ and $2 \times 10^{-8}$). 

The last columns give the ranges of maximum distance of GRO~J1655-40 computed from the constraint that $a \geq 0$, using the lower and upper limits of the absolute magnitude of $M_1$. We note that an {\it HIPPARCOS} parallax error of 1~mas translates into an uncertainty of roughly $\pm$3~pc for the distance $D_2$. The source of uncertainty on the distance of GRO~J1655-40 is therefore dominated by that on the flux ratio $f$. An uncertainty of 20\% on the flux ratio (taking into account 10\% for each of the flux of GRO~J1655-40 and the comparison star) implies an uncertainty of roughly $\pm$ 0.2~kpc on the distance $D_1$.

\begin{table*}[!ht]
\centering
\caption{The five F stars used to compute the maximum distance of GRO~J1655-40 are summarized. The star's name, its spectral type, distance (computed from the {\it HIPPARCOS} parallax), absolute magnitude, ($B-V$) color (obtained from SIMBAD) and the ratio of the calibrated spectrum of GRO~J1655-40 with that of the star are indicated. The last two columns indicate the ranges of maximum distance $D_1$ of GRO~J1655-40 obtained through the constraint of $a \geq 0$ with the two limits of the absolute magnitude $M_1$ and for both methods. See text for details. The uncertainty on the distance values is $\pm$0.2~kpc. The error on the absolute magnitude is obtained from the error on the HIPPARCOS distance. }
\begin{tabular}{llcclrll} \hline 
Star&Sp. Type&Distance&$M$&$B-V$&$f$&Max. $D_1$ (spec)&Max. $D_1$ (phot) \\ 
     &               & (pc)     & (mag) & (mag) &    & (kpc) & (kpc)    \\ \hline
\object{HD~156098} & F6IV  & 50$\pm$2     & 2.0$\pm$0.2 & 0.50 & $( 8.5 \pm 1.2)\,10^{-4}$ & 1.08 - 2.71 & 1.30 - 3.27\\
\object{HD~16673}  & F5V   & 21.5$\pm$0.5 & 4.1$\pm$0.1 & 0.52 & $( 9.3 \pm 1.3)\,10^{-4}$ & 0.46 - 1.15 & 0.51 - 1.27\\
\object{HD~210848} & F7II  & 68$\pm$4     & 1.4$\pm$0.3 & 0.50 & $(11.9 \pm 1.7)\,10^{-4}$ & 1.25 - 3.13 & 1.70 - 4.27\\
\object{HD~37495}  & F4V   & 42$\pm$1     & 2.2$\pm$0.1 & 0.50 & $(10.2 \pm 1.1)\,10^{-4}$ & 0.83 - 2.08 & 1.21 - 3.05\\
\object{HD~65925}  & F5III & 58$\pm$2     & 1.4$\pm$0.2 & 0.40 & $( 8.4 \pm 0.9)\,10^{-4}$ & 1.26 - 3.16 & 1.48 - 3.72\\ \hline
\end{tabular}
\label{multistar}
\end{table*}

In Table \ref{multistar}, none of the upper values of the spectroscopic ranges of maximum distance exceeds 3.2~kpc. Interestingly, using the absolute magnitude $M_1=0.7$ given by \citet{Orosz-Bailyn-1997}, Eq.~\ref{absorption_vs_distance} implies a distance ranging from 0.15 to 1.43~kpc for the five F stars. Apart from the range obtained with the star HD~16673 (F5V), the overall mean maximum distance, that is for obtaining a {\it null} absorption ($a \equiv 0$), is 1.7~kpc (i.e. equivalent to the distance at null absorption for the star HD~156098 only, which has the same spectral type as the secondary star in GRO~J1655-40). We thus consider this value as a strong upper limit to the distance of GRO~J1655-40, since the absorption is certainly not strictly zero. We emphasize here that the only assumption we made is the absence of extinction toward the comparison star (but even a small absorption toward the sources would not challenge our conclusion), and that the absolute magnitude of the F6IV star in GRO~J1655-40 is that of HD156098 within two magnitudes. 

To strengthen these results, we used published photometric data of the comparison stars and applied the pair method. More precisely, we can estimate a $B$ magnitude of GRO~J1655-40 in quiescence of $\sim$16.65 from the lightcurves published by \citet{Orosz-Bailyn-1997}. Using their $V$ magnitude (17.12), we obtain $(B-V)_{GRO} = 1.53$. On the other hand, $(B-V)$ colors of the comparison $F$ stars can be retrieved from SIMBAD, and assuming that there are unreddened because of their close distance (hence $(B-V)_{F*}=(B-V)_{F*,0}$), we can use the following equations \citep[vol II, p33.]{Olson-1975,Lang-1999}: 

\begin{equation}
R = 3.25 + 0.25 \cdot (B-V)_{F*,0} + 0.05 \cdot E(B-V)
\label{R}
\end{equation}
\begin{equation}
A_V = R \cdot E(B-V)
\end{equation}
where $E(B-V) = (B-V)_{GRO} - (B-V)_{F*,0}$. Using the same assumption about the absolute magnitude of the secondary star of GRO~J1655-40, ranges of maximum distance can be computed using the standard equation: $V-M_V-A_V = 5 \log D -5$. These ranges are given in the last column of Table~\ref{multistar}. It can be seen that they agree with the ranges given by the spectral flux method, although the mean distance (excepting HD~16673) is slightly larger: 2.25~kpc.

\section{Discussion}

No astrophysical link has been invoked in the literature between GRO~J1655-40 and its environment, except \citet{Mirabel-etal-2002} who mention the possible link of the source with the open cluster NGC~6242, since the opposite direction of the proper motion vector of GRO~J1655-40 that they determined with {\it HST} is clearly pointing to the cluster. This open cluster lies at a distance of $1.02 \pm 0.1$~kpc from the Sun according to the photometric and radial-velocity measurements of \citet{Glushkova-etal-1997}. Although the association of  GRO~J1655-40 with the open cluster cannot be proven with our data, the upper limit for the distance of GRO~J1655-40 is fully consistent with the distance of the cluster. Assuming that the link is true, using the proper motion values published by \citet{Mirabel-etal-2002} and the distance between the source and the cluster NGC~6242, we can compute the time at which GRO~J1655-40 would have been ejected from the star cluster : $\sim6.6 \times 10^5$ years ago. Since the actual age of the cluster is known: 40.55 $\times~10^6$ years \citep{Kharchenko-etal-2005}, we can deduce that the age of NGC~6242 when the progenitor of GRO~J1655-40 exploded is roughly 40 millions years. Using the canonical isochrones of \citet{Pietrinferni-etal-2004} for a solar metallicity\footnote{{\tt http://www.oa-teramo.inaf.it/BASTI/}}, it is possible to obtain the turn-off mass of a cluster with the given age: between 6.7 and 7.2 $M_{\odot}$ (assuming that the progenitor of the black hole in GRO~J1655-40 has followed a single-star evolution). If correct, this result favors the lower end of the mass range given by \citet{Shahbaz-etal-1999}, and is incompatible with the values obtained by \citet{Orosz-Bailyn-1997}. Finally, using the radial velocity of the system \citep[$-142.4\pm$1.6 km~s$^{-1}$][]{Orosz-Bailyn-1997}, and assuming the age given above, we note that GRO~J1655-40 is closer to the Sun than NGC~6242 by about 100~pc.

\section{Conclusion}

We have shown that the distance of GRO~J1655-40 of 3.2~kpc quoted in the literature has been obtained through a refinement of a distance range, which was in turn not well established. We have determined a spectral type F6IV for the secondary star using UVES spectra. By flux-calibrating our spectra of GRO~J1655-40 during quiescence and comparing them with various stars of similar spectral types, we have shown that the distance of GRO~J1655-40 simply cannot be of 3.2~kpc, and is certainly smaller than 1.7~kpc. We consider the possibility that the source is associated with the open cluster NGC~6242 \citep[see Fig.~1 of][]{Mirabel-etal-2002}, which lies at a distance of 1.0~kpc. With such distance, GRO~J1655-40 is not a superluminal source (the jets speed reaches $\sim$ 0.37$c$). If the distance is confirmed (maybe by the future european satellite {\it Gaia}), and assuming that the distance of other black-hole binaries are correct, GRO~J1655-40 would become one of the closest known black hole to the Sun.

\begin{acknowledgements}
The authors thank L. Schmidtobreick for very useful discussions, and for pointing out an important weakness in our argumentation on the absorption. We thank the anonymous referee for constructive remarks. C.F. wants to also thank C. Sterken for useful comments. This research has made use of the SIMBAD database, operated at CDS, Strasbourg, France. The authors warmly thank Gijs Nelemans for noticing an error in our computation of distance ranges with the photometric method.
\end{acknowledgements}

\bibliographystyle{aa}

\end{document}